\documentclass{cupconf}

\usepackage{graphicx}

  \checkfont{eurm10}
  \iffontfound
    \IfFileExists{upmath.sty}
      {\typeout{^^JFound AMS Euler Roman fonts on the system,
                   using the 'upmath' package.^^J}%
       \usepackage{upmath}}
      {\typeout{^^JFound AMS Euler Roman fonts on the system, but you
                   dont seem to have the}%
       \typeout{'upmath' package installed. cupconf.cls can take advantage
                 of these fonts,^^Jif you use 'upmath' package.^^J}%
      }
  \else
  \fi

  \checkfont{msam10}
  \iffontfound
    \IfFileExists{amssymb.sty}
      {\typeout{^^JFound AMS Symbol fonts on the system, using the
                'amssymb' package.^^J}%
       \usepackage{amssymb}%

      }{}
  \fi

  \IfFileExists{amsbsy.sty}
    {\typeout{^^JFound the 'amsbsy' package on the system, using it.^^J}%
     \usepackage{amsbsy}}
    {\providecommand\boldsymbol[1]{\mbox{\boldmath $##1$}}}

\newsavebox{\astrutbox}
\sbox{\astrutbox}{\rule[-5pt]{0pt}{20pt}}

\newcommand\etal{\mbox{\textit{et al.}}}

\title[Evolution of SMBHs]{Evolution of supermassive black holes}

\author[M\"uller \& Hasinger]%
{A\ls N\ls D\ls R\ls E\ls A\ls S\ns M\ls {\"U}\ls L\ls L\ls E\ls R\break
\and G\ls {\"U}\ls N\ls T\ls H\ls E\ls R\ns H\ls A\ls S\ls I\ls N\ls G\ls E\ls R}

\affiliation{Max--Planck--Institut f\"ur Extraterrestrische Physik,  PO box 1312, 85741 Garching, Germany}

\pubyear{2007}
\volume{???}
\pagerange{???--???}
\date{?? and in revised form ??}
\begin{document}

\maketitle

\begin{abstract}
The cosmological evolution of supermassive black holes (SMBHs) seems to be intimately linked to  
their host galaxies. Active galactic nuclei (AGN) can be probed by deep X--ray 
surveys. We review results from large X--ray selected samples including first 
results from the XMM--Newton COSMOS survey. A new picture arises from the fact that high--luminosity 
AGN grow earlier than low--luminosity AGN. In particular, the space density of low--luminosity AGN 
exhibits a significant decline for redshifts above z=1. This ''anti--hierarchical'' growth scenario of 
SMBHs can be explained by two modes of accretion with different efficiency. 
The population of Compton--thick sources plays a key role in our understanding of the BH growth 
history. Their space density and redshift distribution is relevant to estimate the SMBH mass function. 
A comparison with the relic SMBH mass distribution in the local Universe constrains the average 
radiative efficiency and Eddington ratio of the accretion. 
We discuss a new synthesis model of Compton--thin and Compton--thick sources that is in 
concordance with deep X--ray observations and in particular predicts the right level of contribution
of the Compton--thick source population observed in the Chandra Deep Field South observations as well 
as the first INTEGRAL and Swift catalogues of AGN. Currently, one of 
the most important problems is the evolution of obscuration with redshift.
\end{abstract}

\firstsection 
\section{Introduction}

\subsection{Deep surveys}
Deep field observations are a suitable observational technique to probe AGN physics. In multi--wavelength campaigns, 
astronomers select a field in the sky and produce images from several pointings. They use a combination of wide fields 
to collect many sources and so--called pencil beams to investigate sources in deep space. Today, there are many samples 
in different wavelength bands available and large surveys are underway. Among the most important deep extragalactic surveys to date are the two fields related with the GOODS survey, the Hubble Deep Field North (HDF—N) and the Chandra Deep Field South (CDF--S), the latter field encompassing the Hubble Space Telescope ACS Ultradeep Field (HUDF) and embedded in the larger GEMS, COMBO-17 and ECDFS surveys. The COSMOS project is a multiwavelength survey centered around the largest HST observing project, covering 2 square degrees.   

On the basis of multi--wavelength observations and in comparison with galaxy and accretion models it turns out that AGN are 
mainly determined by the following physical parameters: accretion rate, accretion efficiency, SMBH mass, SMBH spin, stellar 
formation rate, metallicity, dusty torus mass, and seed magnetic field. 
One geometric parameter is the inclination angle of the system. The challenge consists in determining these parameters from 
observations and to fit them into an AGN unification model.

The innermost part of AGN can be probed by means of X--ray observations. The hot inner accretion flows produce soft and hard 
X--rays. Hence, X--ray deep surveys provide an important issue to study AGN and their SMBHs inside.

Thereby, astronomers are confronted with observational challenges as well. In X--ray deep surveys the redshift of the sources 
is optically determined by using spectroscopic techniques (spectro--z) or by photometric techniques (photo--z). At high cosmological 
redshifts the I band magnitudes are faint and in certain redshift ranges spectroscopic features are weak or absent, therefore astronomers fail to determine spectro--z's. The problem of spectroscopic 
incompleteness emerges typically at $1<z<2$ -- a phenomenon that was coined 'spectroscopic desert' (see e.g. \cite{Brandt2005}). 
%
%
\begin{figure}
\begin{center}
\includegraphics[width=\textwidth]{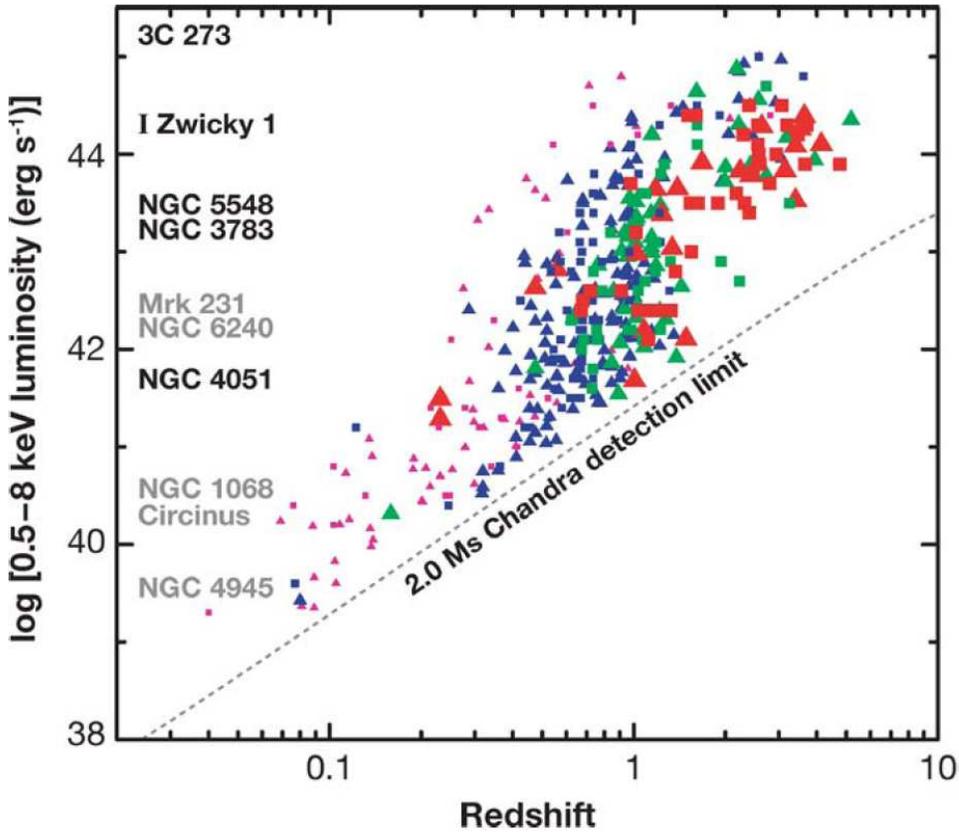}
\caption{Rest--frame 0.5-8 keV luminosity as function of redshift for sources form the CDF--N ({\em triangles}) and CDF--S 
({\em squares}). The data points are grouped above the 2 Ms Chandra detection limit ({\em dotted line}). The 'spectroscopic 
desert' is visible on the right--hand side in the luminosity--redshift plane (taken from \cite{Brandt2005}).}\label{fig:desert}
\end{center}
\end{figure}
Figure~\ref{fig:desert} illustrates the sources in the luminosity--redshift plane. The I band magnitudes successively 
grow from left to right until the emission drops below the spectroscopic sensitivity limit of 8--10 m class telescopes, 
leaving a lack of spectroscopically identified sources on the on the right--hand side of the plot close to the sensitivity limit. In the CDF--S it is possible to fill this spectroscopic desert with photo--z thanks to a combination of the deep multiwavelength photometry available through GOODS, GEMS and COMBO-17 in this field (\cite{Szokoly2004, Zheng2004, Mainieri2005, Wolf2004, Grazian2006}). As a result, it is possible to obtain 95\% redshift completeness for the CDF--S. Redshift completeness of X--ray flux limited samples is vital to produce statistically and systematically reliable results. Therefore in the further analysis we typically cut all samples to a $>80\%$ redshift completeness, leading to an overall completeness of the sample of $\sim90\%$. 

\subsection{AGN classification criteria}
The classical discrimination between type--1 (unobscured) and type--2 (obscured) AGN is done using a classification through optical spectroscopy. Optical type-1 AGN have broad permitted emission lines, while optical type--2 AGN do not show broad permitted lines, but still have high-excitation narrow emission lines. Therefore, many works in the field are mainly using the presence of broad lines as discriminator and classify type-1 AGN only as Broad-line AGN (BLAGN). However, this AGN classification scheme breaks down, when the optical spectrum is insufficient to discriminate the line widths. At high redshifts and low luminosities there are several effects that compromise the optical classification. First, in general high redshift objects are faint, so that they require very long observing times to obtain spectra of sufficient quality and unambiguously discern broad emission line components. Secondly, there are redshift ranges, where the classical broad lines shift out of the observed optical bands (see above). Thirdly, and probably most importantly, at high redshifts the spectroscopic slit is filled with the whole light of the AGN nucleus and the host galaxy. Depending on the ratio between nuclear and host luminosity, the host galaxy can easily outshine the AGN nucleus, rendering the AGN invisible in the optical light. This is the main reason, why X--ray samples are much more efficient in picking up low-luminosity AGN at high redshift.

In principle, a classification purely based on X--ray properties would be possible. Large hydrogen column densities block soft X--rays. Therefore, a suitable AGN classification scheme in X--rays could involve 
the column density $N_{\rm H}$ that can be determined from the X--ray spectra. Usually, AGN X--ray spectra are fit by simple power law models. 
The effect of increasing column density is that the soft part of the spectrum is more and more suppressed, i.e.\
the spectrum becomes harder at high $N_{\rm H}$. Correlations between optical and X--ray properties of obscured and unobscured AGN show a very good match ($\sim 90\%$) between optical obscuration and X--ray absorption. It is therefore suitable to define an X--ray type--1 AGN by 
$\log N_{\rm H} <  22$ whereas AGN type--2 satisfy $\log N_{\rm H} > 22$\footnote{Luminosities and column densities are 
given in the usual units erg/s and cm$^{-2}$ respectively.}. However, in practice is not possible to determine $N_{\rm H}$ values for samples with typically very faint X--ray sources, due to the small number of observed photons. Therefore, one has to resort to hardness ratios measured from coarse X--ray bands. The determination of $N_{\rm H}$ values from X--ray hardness ratios, in particular at high redshifts is biased towards too high absorption values (see e.g. \cite{Tozzi2006}), so that the X--ray classification alone also introduces significant systematic errors.

To overcome the difficulties of classification in either the optical or the X--ray bands, a combined optical/X--ray classification scheme has been introduced (\cite{Szokoly2004, Zheng2004}). It involves a threshold X--ray luminosity $L_{\rm X}$ 
and the X--ray hardness ratio HR, defined as 
\begin{equation}
{\rm HR} = ({\rm H}-{\rm S}) / ({\rm H}+{\rm S}),
\end{equation}
with count rate in the hard band H and soft band S. The threshold AGN X--ray luminosity is set at for an AGN $\log L_{\rm X}>42$, roughly at the upper luminosity range of vigorously star forming galaxies. The practical combined optical/X--ray AGN classification is the following: An optical broad--line AGN (BLAGN) or a galaxy with $\log L_{\rm X} > 42$ and hardness ratio ${\rm HR} < -0.2$ is called an AGN type--1. An optical narrow-line AGN (NLAGN) or a galaxy with $\log L_{\rm X} > 42$ and hardness ratio ${\rm HR} > -0.2$ is called an AGN type--2. 

This classification has the advantage that it can also be applied for sources, which have only photometric redshifts and no spectroscopic information at all. The other advantage is, that it is based on a simple measurable quantity, the hardness ratio. There are also obvious systematic effects and difficulties. A small fraction of AGN are known not to follow the simple optical/X--ray obscuration/absorption correlation, either broad line AGN have absorbed X--ray spectra (e.g. for BAL QSOs), or narrow-line AGN have unabsorbed X--ray spectra (e.g. for narrow-line Seyfert-1 galaxies or Compton-thick, reflection dominated spectra). Also the hardness ratio threshold is a function of redshift. Finally, the dust obscuration versus gas absorption properties may be different, depending on environment and redshift. Nevertheless, we consider these systematic effects far smaller than those of any of the other classifications individually. All the systematics addressed here can in principle be quantified through appropriate theoretical population synthesis models.

\section{Separate evolution of AGN classes}
%
\subsection{Anti--hierarchical growth of SMBHs}
Deep surveys supply the X--ray luminosity functions (XLF) and bolometric luminosity functions of AGN. Using soft X--ray luminosity 
functions, Hasinger \etal (2005) 
could confirm that the density evolution depends significantly on AGN luminosity, i.e.\ the cosmological evolution of high--luminosity 
AGN (HLAGN) deviates strongly from that of low--luminosity AGN (LLAGN): the space density of HLAGN peaks at higher redshift $z\sim 2$
but the space density of LLAGN peaks at $z\sim 0.7$. This systematic trend of active SMBHs has been coined 'anti--hierarchical growth' or 'cosmic downsizing'. It is 
possible to understand this behaviour in the framework of accretion theory (\cite{Merloni2004}). The observed local black hole mass 
function (BHMF) 
is taken as boundary condition to integrate the continuity equation backwards in time (mergers are neglected). It could be shown that
the most massive SMBH with billions of solar masses were already in place at $z\simeq 3$. Indeed, this is observed for some SDSS quasars 
\cite{Fan2003}. 
Merloni (2004) has also demonstrated that radiatively inefficient accretion (e.g.\ ADAF solutions) dominate only at $z < 1$. Therefore,
Seyfert galaxies and quasars evolve cosmologically in a radically different manner. The bulk of AGN has to wait much longer to grow 
or to be activated.  

%
\begin{figure}
\begin{center}
\includegraphics[width=\textwidth]{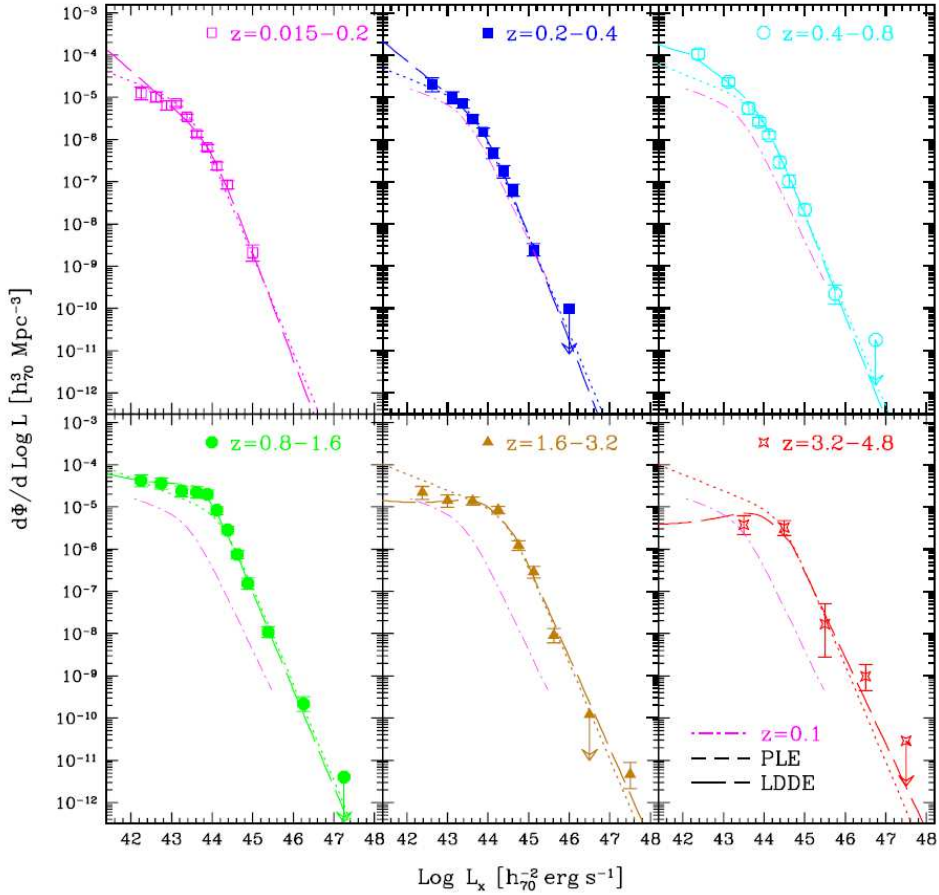}
\caption{Observed 0.5-2 keV luminosity functions for AGN type--1 in six different redshift ranges. The solid curves through the data 
points are best luminosity--dependent density evolution (LDDE) fits. For comparison, the curve from the upper left panel ($z=0.015-0.2$) 
is plotted in all other panels. Adapted from \cite{Hasinger2005}.}\label{fig:ldde}
\end{center}
\end{figure}
The soft X--ray (0.5-2 keV) luminosity functions for AGN type--1 in six different redshift ranges are shown in figure~\ref{fig:ldde}. 
Here, a successive evolution along redshift is visible. Only a luminosity--dependent density evolution (LDDE) can describe the data.
Pure luminosity evolution (PLE) or pure density evolution (PDE) models are inadequate.

Now, there is firm evidence of a decline in space density of LLAGN towards higher redshift. For HLAGN, i.e.\ high--mass BHs deep X--ray surveys so far provide rather poor constraints about their growth phase in the redshift range $3\lesssim z\lesssim 6$ (\cite{Hasinger2005, Silverman2005}), while 
in the optical and radio bands a clear decline of space density is observed for $ z>3$ (\cite{Fan2001, Wall2005}).

The observed luminosity functions have also been compared to other models. \cite{Marconi2004} 
compared the mass function of black holes (BHs) in the local universe with that from AGN relics. The basic idea is that the relic BHs are 
grown from seed BHs by accretion in AGN phases. The local black hole mass function (BHMF) is estimated by using the prominent correlations between
BH mass, stellar velocity dispersion and bulge luminosity. Again, merging is neglected in this model. Then, it is possible to derive the 
average radiative efficiency $\epsilon$ and the ratio between emitted and Eddington luminosity $\lambda=L/L_{\rm Edd}$ and the average 
lifetime of active BHs. Based on this model, typical efficiencies are found to be $\epsilon=0.04-0.16$ and Eddington luminosities 
$\lambda =0.1-1.7$. The BH lifetime strongly depends on BH mass and amounts to 450 million years for 100 solar mass BHs and 150 million 
years for most massive BHs with more than billion solar masses.

The luminosity functions from SDSS (\cite{Richards2006}),  
from hard X--rays (\cite{LaFranca2005}) 
and soft X--rays (Hasinger \etal~2005) 
are fairly consistent with the model by Marconi \etal (2004)\footnote{even if bolometric corrections are not applied}. Soltan's 
original argument
(\cite{Soltan1982}) can be extended: one compares the accreted BHMF derived from the cosmic X--ray background with the MF of dormant relic 
BHs in local galaxies. Both can be reconciled if an energy conversion efficiency of $\epsilon=0.1$ is assumed. From accretion theory it 
is known that such high efficiencies involve a Kerr BH, i.e.\ active BHs in AGN have to rotate. Other work on cosmological BH growth also 
indicates high BH spins (\cite{Volonteri2005}).  

Recently, bolometric quasar luminosity functions have been determined and they agree well with the X--ray data in each redshift slice.
They demonstrate a perfect match between the soft X--ray, hard X--ray, optical and even MIR wavebands, when the appropriate bolometric corrections and absorption distributions are taken into account \cite{Hopkins2007}. 
Because the optical and soft X--ray luminosity functions refer to type-1 AGN only, while the hard X--ray and MIR luminosity functions refer to all AGN, this agreement gives a kind of sanity-check on the type--1/type--2 classification discussed above. \cite{Barger2005}, using a pure BLAGN classification, on the contrary find a strong decline of the type--1 space density at the low--luminosity end. 

Recent multi--scale simulations have shown that it is possible to explain the highest redshift, massive SDSS quasar SMBHs by gas accretion plus major mergers
in a small group of protogalaxies within the framework of standard $\Lambda$CDM cosmology (\cite{Li2006}).
This picture immediately calls for deeper observations because the known quasars with billion solar masses SMBHs at $z\simeq 6$ should possess
progenitors, i.e.\ the so--called mini-quasars with million solar masses at $z\simeq 10$ and stellar BHs with a few hundred solar masses at 
$z\simeq 20$. It is the task of future X--ray missions to probe these progenitors.

\subsection{Luminosity--dependence of AGN type--2 fraction}
\begin{figure}
\begin{center}
\includegraphics[width=\textwidth]{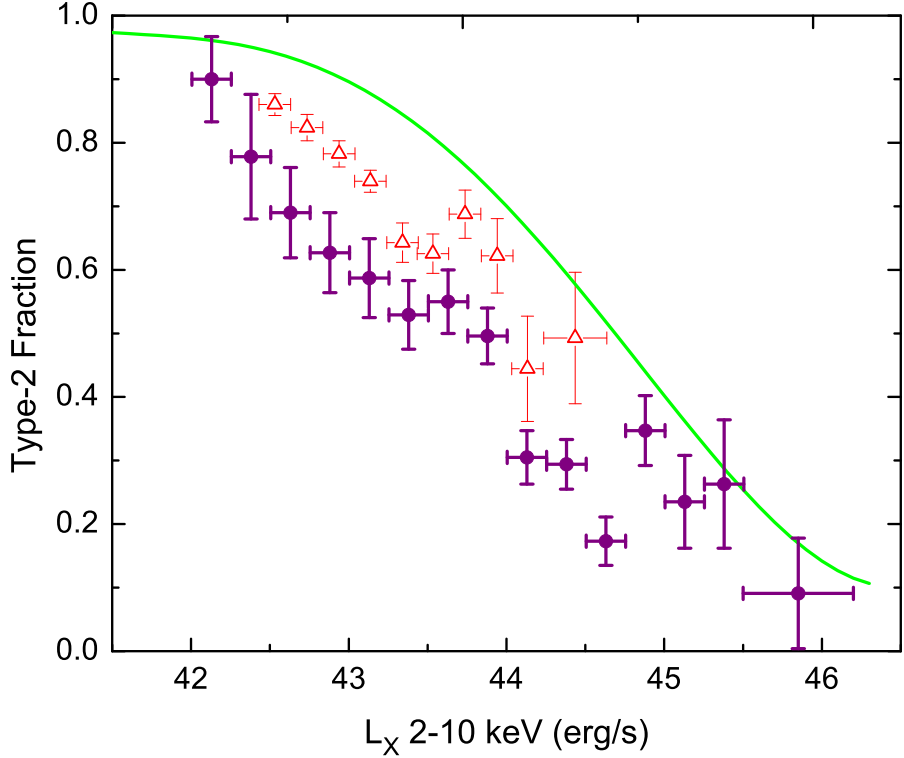}
\caption{The luminosity dependence of the type--2 fraction compiled from different work. The X--ray data (filled circles) are from a sample of 1029 AGN selected in the 2-10 keV band, with $>90\%$ redshift completeness (\cite{Hasinger2007}). The triangles are from a sample of [OIII] selected AGN from the SDSS \cite{Simpson2005}. The solid line is from silicate dust studies of a sample of AGN observed with Spitzer in the MIR (\cite{Maiolino2007}). Given the known selection effects in the different bands, the data are consistent with each other and exhibit the same trend of decreasing absorption with luminosity.}\label{fig:fraction}
\end{center}
\end{figure}

A grouping into luminosity classes is one issue to prove the separate evolution of AGN classes. Here we summarize recent work on the 
AGN type--2 fraction. \cite{Ueda2003} studied the hard X--ray luminosity functions of an ASCA sample with $\sim$230 sources which 
was highly complete, $\sim 95$\%. They found that the fraction of type--2 AGN decreases with X--ray luminosity, $L_{\rm 2-10\,keV}$. Similar results have been obtained with different systematics by the CLASXS team (\cite{Steffen2003}) and by Hasinger (2004). 
Our preliminary analysis (\cite{Hasinger2007}) confirms this trend in much more detail and better statistics using different data sets from XMM--Newton, Chandra, ASCA and HEAO-1 (see figure \ref{fig:fraction}). 
Further, [OIII]--selected Seyfert galaxies of the SDSS agree with this trend \cite{Simpson2005}. 
It has been demonstrated in this work that for Seyfert galaxies selected in the SDSS, the fraction of BLAGN increases with the luminosity of the isotropically emitted [O III] narrow
emission line. The [O III] luminosity can be converted into a 2-10 keV luminosity by empirical relations found for Seyfert galaxies 
\cite{Mulchaey1994}, i.e.\ $L_{{\rm[O III]}}=0.015\,L_{\rm 2-10\,keV}$. 
  
Figure~\ref{fig:fraction} illustrates all this work in direct comparison.
%
%
Of course, both the X--ray and optical wavelength bands are subject to selection effects and biases: X--rays in 
the 2--10 keV band miss the population of Compton--thick Seyferts; optical surveys miss the AGN population diluted by the host galaxy. 
Nevertheless, the agreement of better than 20\% between the optical and X--ray selection is reassuring. 
Physically, this luminosity--dependent fraction might be interpreted as a 'cleanout effect' because more luminous AGN can dissociate, ionize and finally  blow away the dust in their environment. 
Recently, mid--infrared (MIR) Spitzer observations of 25 luminous and distant quasars additionally confirm this picture (\cite{Maiolino2007}). 
They find a non--linear correlation between 6.7$\mu$m MIR emission and 5100 \AA~optical emission. This is interpreted as decreasing covering 
factor of circumnuclear dust as function of luminosity. Further, the silicate emission strength correlates with the luminosity, the accretion 
rate $L/L_{\rm Edd}$ and black hole mass $M_{\rm BH}$.

The evolution of this luminosity--dependent obscuration fraction with redshift is still a matter of intense debate. A redshift depencence of the obscured fraction was claimed by \cite{Fiore2003}, using the HELLAS2XMM sample. On the other hand, other authors (\cite{Ueda2003, Gilli2007}) did not find a redshift dependence. Recently, Treister and Urry (2006) 
performed a similar analysis, using an AGN meta--sample of 2300 AGN with $\sim50\%$ redshift completeness and find a shallow increase of the  
type--2 AGN fraction proportional to $(1+z)^{0.3}$. The differences between the various results could be due to the way the redshift incompleteness, the different flux limits and the AGN classification have been taken into account in the analysis. 

%
\begin{figure}
\begin{center}
\includegraphics[width=\textwidth]{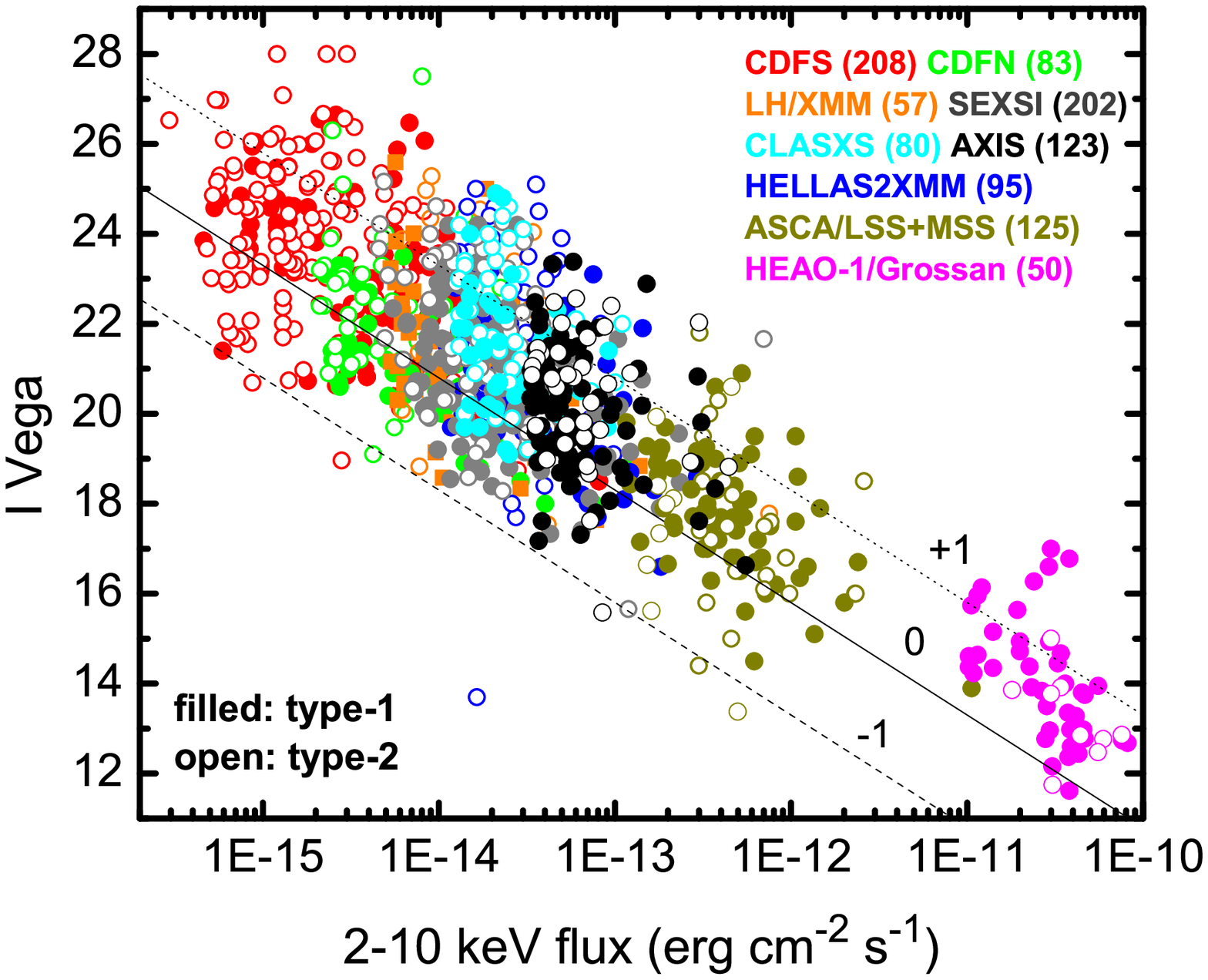}
\caption{Hard X--ray (2--10 keV) sample of 1029 AGN in the I--flux plane. The individual samples have been cut a flux limit yielding $>80\%$ redshift completeness: CDF--S (\cite{Szokoly2004, Zheng2004, Grazian2006}), CDF--N (\cite{Barger2003}), LH/XMM (\cite{Mainieri2002, Szokoly2007}), SEXSI (\cite{Eckart2006}), CLASXS (\cite{Steffen2004}), AXIS (\cite{Barcons2007}), HELLAS2XMM (\cite{Fiore2003}), ASCA LSS \& MSS (\cite{Akiyama2000, Ueda2005}), HEAO-1 / Grossan (\cite{Shinozaki2006})}\label{fig:hsam1}
\end{center}
\end{figure}
%
%
\begin{figure}
\begin{center}
\includegraphics[width=\textwidth]{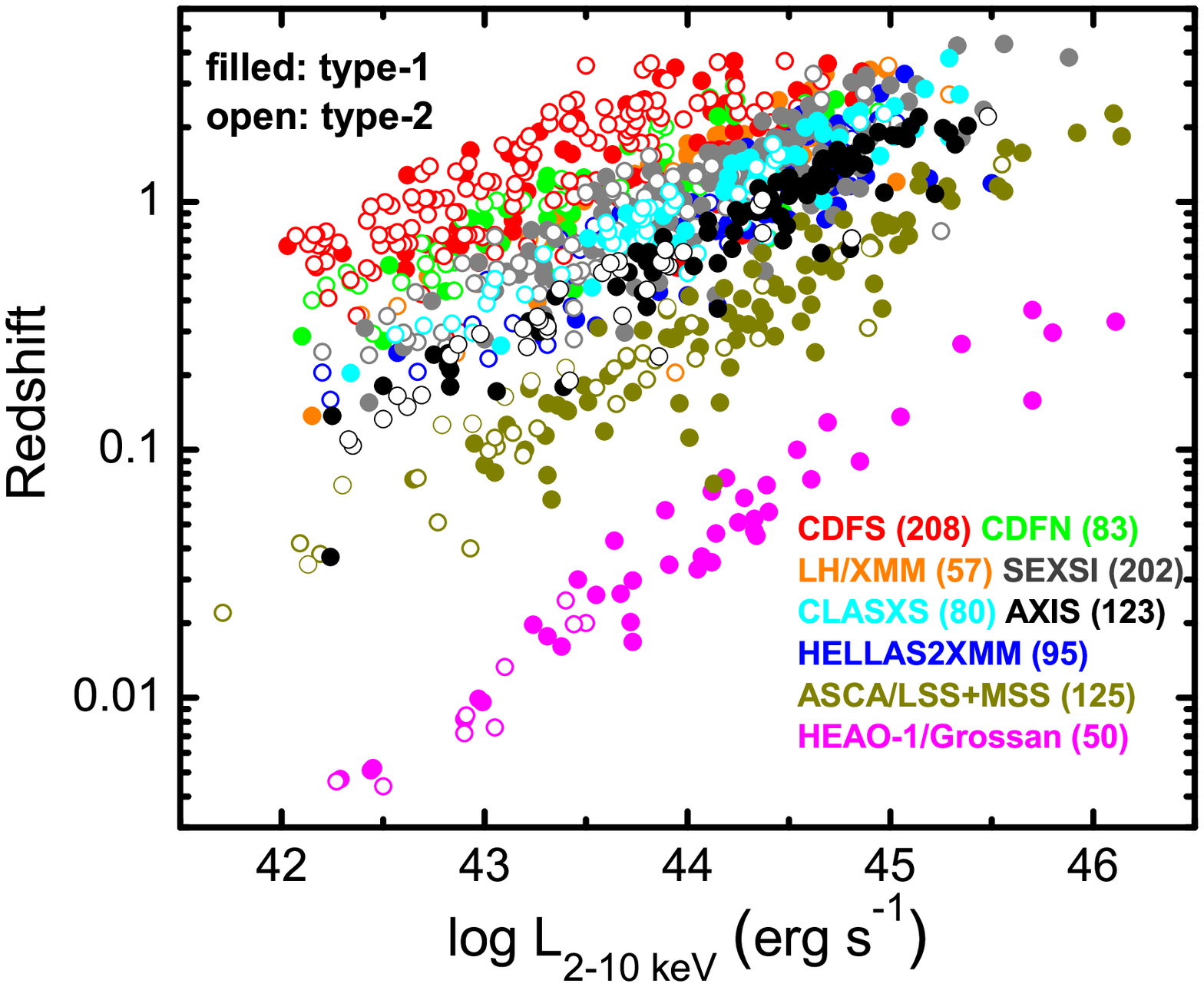}
\caption{The same sample as figure \ref{fig:hsam1} in the redshift--luminosity plane.}\label{fig:hsam2}
\end{center}
\end{figure}

Here, we shortly report on ongoing work (\cite{Hasinger2007}) on a sample of $\sim$1000 AGN selected in the 2--10 keV band with $\sim$90\% optical redshift completeness. 
In figure \ref{fig:hsam1} and \ref{fig:hsam2} we show the 2-10 keV AGN samples utilized in this analysis in an X--ray flux/optical magnitude and an X--ray luminosity/redshift diagram. Type--1 and type--2 AGN of the different samples are shown with filled and open circles, respectively. Just looking at the data, there is a noticable change of the fraction of type--2 AGN as a function of luminosity. This trend becomes very obvious, if the data is integrated in the redshift range z=0.2--3.2 and the type--2 fraction is displayed as a function of X--ray luminosity (figure \ref{fig:fraction}.) 

\begin{figure}
\begin{center}
\includegraphics[width=\textwidth]{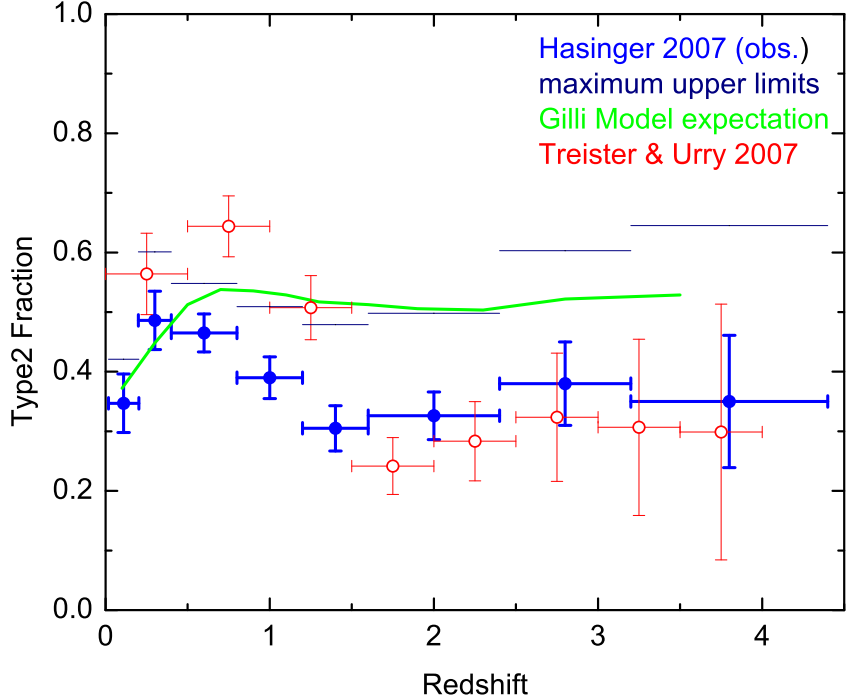}
\caption{The observed fraction of type--2 AGN as a function of redshift. Filled Symbols and horizontal upper limits are from \cite{Hasinger2007}, open symbols from \cite{Treister2006}. The solid line shows the prediction of the \cite{Gilli2007} model, assuming no obscuration evolution (courtesy: Roberto Gilli).}\label{fig:evolution}
\end{center}
\end{figure}

In order to address a possible evolution of the obscuration fraction with redshift, one can first simply determine the observed ratio of type--2 versus total AGN, integrated over all luminosities, in shells of increasing redshift. This is very similar to what was done by Treister \& Urry (2006). Figure \ref{fig:evolution} shows the observed change of the observed fraction of absorbed objects with redshift. Despite the rather different selection effect and redshift incompleteness (90\% vs. 50\%), 
the observed data of Hasinger (2007) and Treister \& Urry (2006) agree very well at redshifts above $z=1.5$. At lower redshifts the Treister \& Urry values are systematically above those of Hasinger, which is probably due to the fact, that the former authors have only included BLAGN in their type-1 selection and thus systematically overpredict the number of obscured sources. 

The diagram also shows an attempt to estimate the possible effects of the small but still significant redshift incompleteness in the sample. For each redshift shell all the unidentified sources were placed at the center of the corresponding shell and included in the type--1 or type--2 classification, depending on their measured hardness ratios and assumed luminosities. This is clearly a very conservative limit, since it is highly unlikely that all the missing redshifts fall into one bin. Finally, the diagram also shows the type--2 fraction as a function of redshift predicted from the most recent background synthesis model (\cite{Gilli2007}). This curve shows, that the steep rise of the observed type--2 fraction at low redshift can be understood as an effect of the different flux limits for type-1 and type--2 sources in each survey (the latter ones are harder to find because of absorption). The Gilli et al. model predicts a smaller variation with redshift, than what is actually observed. This is probably due to the fact that the model assumes a flat type--2 fraction at high luminosities, while the observations in figure \ref{fig:fraction} indicate a continuing decline. 

%
%

\begin{figure}
\begin{center}
\includegraphics[width=\textwidth]{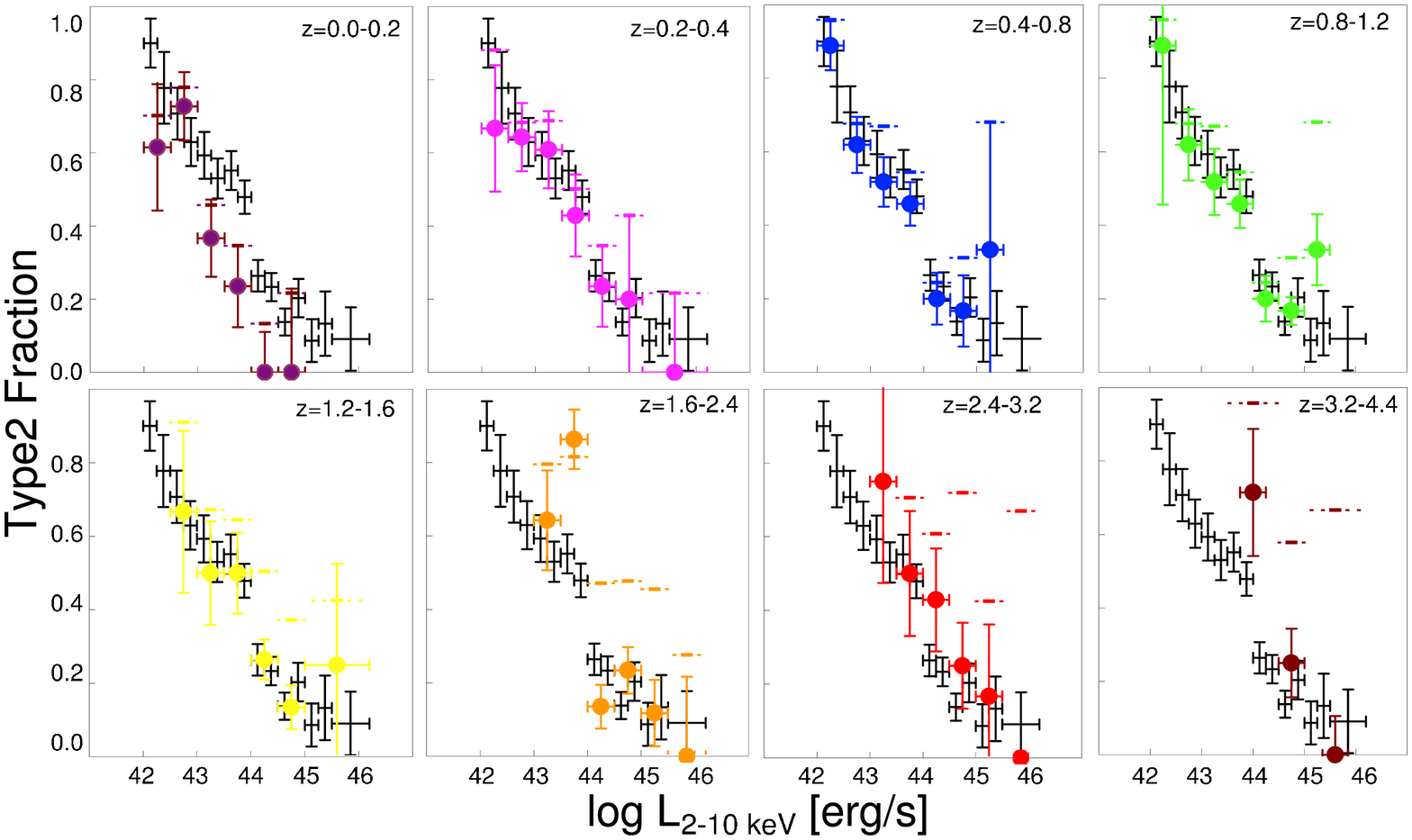}
\caption{Type-2 fraction as a function of X--ray luminosity in different redshift shells. Filled circles show the data in the individual redshift shell indicated on each panel, while crosses give the average relation in the z=0.2-3.2 range for comparison.}\label{fig:lxz}
\end{center}
\end{figure}

Figure \ref{fig:evolution} is integrated over all luminosities in each redshift shell and therefore possibly hiding trends, which are observable in a better resolved parameter space.
In figure \ref{fig:lxz} we therefore finally show the type--2 fraction as a function of luminosity, determined in the different redshift shells and compared to the mean relation in the $z=0.2-3.2$ interval. A similar decreasing type--2 fraction as a function of luminosity is found in each redshift interval. At closer look, there is however a small trend of increasing obscuration fraction with redshift, even ignoring the lowest redshift shell, which is affected most by the flux limit incompleteness (see figure \ref{fig:evolution}). A more quantitative analysis will be shown in Hasinger (2007).

%
%

%

\section{The cosmic X--ray background}
\begin{figure}
\begin{center}
\includegraphics[width=\textwidth]{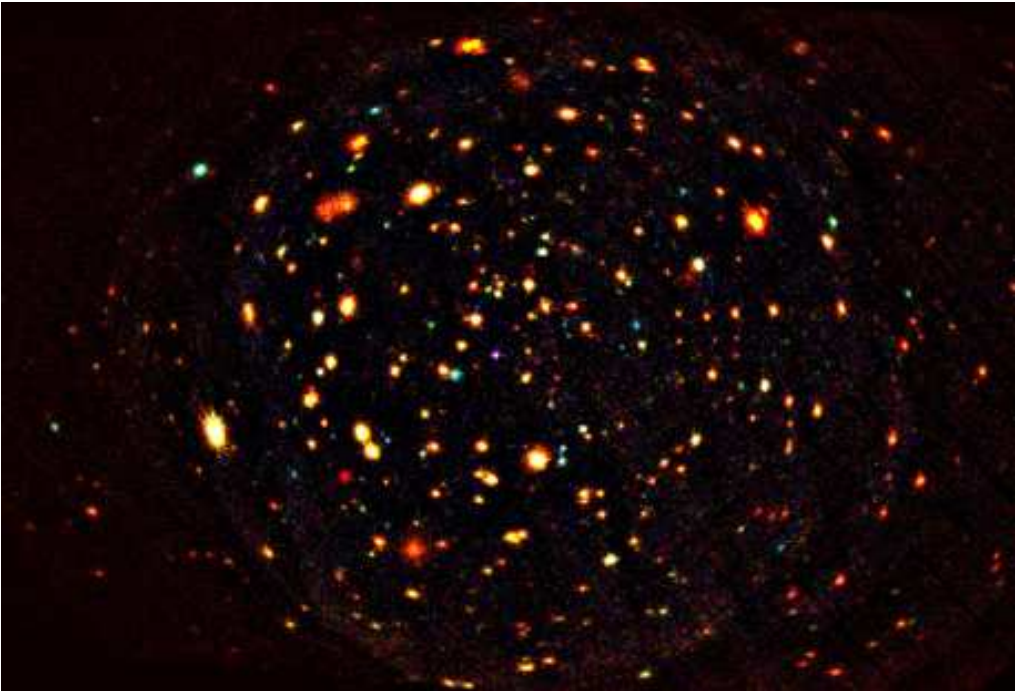}
\caption{X--ray deep field image taken in the Lockman Hole with XMM--Newton (\cite{Hasinger2001}). The total exposure time amounts to $\sim$800 ks. 
This is a X--ray false colour composite image with red, green, and blue corresponding to 0.5-2, 2-4.5 and 4.5-10 keV photon energy. Point--like 
and extend sources are clearly visible that are associated with AGN and galaxy clusters respectively.}\label{fig:lockman}
\end{center}
\end{figure}

One major goal in X--ray astronomy is to explain the origin of the cosmic background emission in detail. The observed emission in X--rays 
originates mainly from AGN (point sources) and galaxy clusters (extended sources) (\cite{Giacconi1962, Hasinger2004, Brandt2005}). 
Figure~\ref{fig:lockman} displays the typical appearance of an XMM--Newton X--ray deep field, located in the Lockman Hole in this case (\cite{Hasinger2001}). 
The Lockman Hole is an extraordinary location at the sky due to the fact that the column density is very low, 
$N_{\rm H}=5.7\times 10^{19} \ {\rm cm}^{-2}$ (\cite{Lockman1986}).
%
%

Different groups developed detailed population synthesis models to reconstruct the observed cosmic X--ray 
background (XRB). Highly obscured AGN turn out to be crucial in this modelling. Even in the local universe, it 
has been shown that $\simeq$40\% of the Seyfert galaxies are Compton--thick (\cite{Risaliti1999}). 
This means that these sources are missing from all current X--ray surveys because their column 
density is extremely large, $\log N_{\rm H}\gtrsim 25$. As a consequence, 
only the reflected emission is seen in these X--ray spectra. This hidden AGN population should show up 
in mid--infrared surveys. The flux criterion $F_{E > 10\,{\rm keV}}\gg F_{\rm SX}$ may serve as 
another good indicator for Compton--thick sources that could be investigated with INTEGRAL or Swift. Recently, \cite{Ueda2007} have presented Suzaku observations of two AGN out of a small sample detected in the Swift BAT survey, which show substantial X--ray absorption and may well be moderately obscured Compton-thick objects. 
Based on Chandra data from the CDF--S, Tozzi \etal (2006) have found $\simeq$5\% of AGN 
are Compton--thick and in the XMM--Newton observations of the COSMOS field a similar fraction of candidate Compton--thick AGN
is found (\cite{Hasinger2006}). 
Beyond $z\simeq 0.1$ Compton--thick AGN are, however, very sparse (\cite{Martinez2006}). 
%
%
\begin{figure}
\begin{center}
\includegraphics[width=\textwidth]{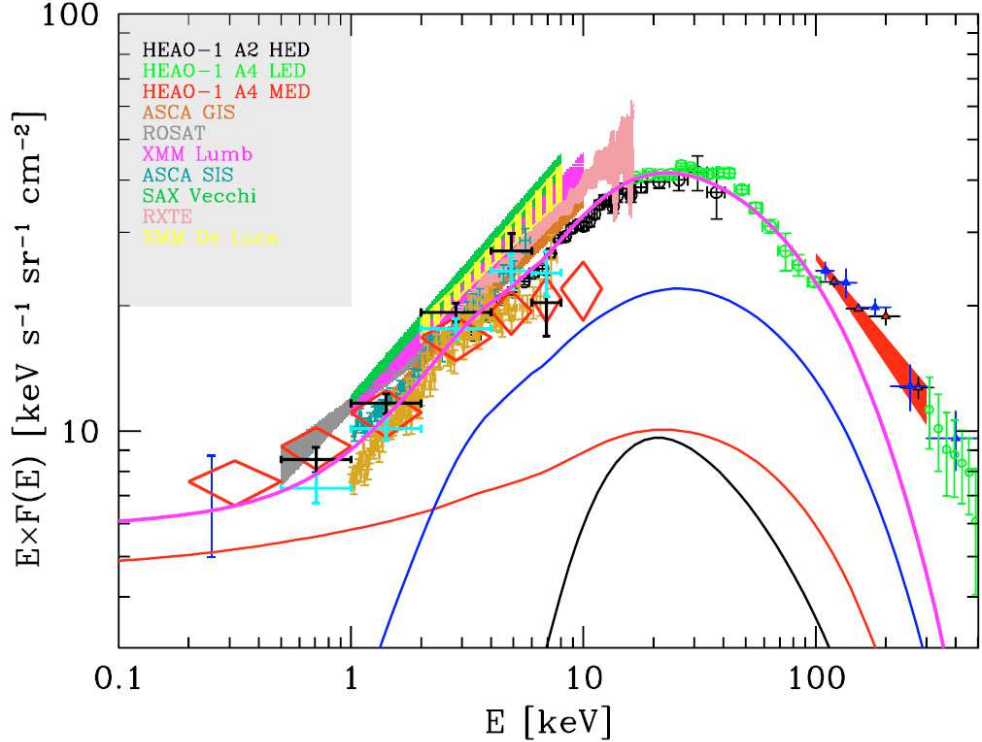}
\caption{The observed cosmic XRB spectrum ({\em data points}) and the predicted contributions of different AGN families ({\em solid lines}) 
taken from \cite{Gilli2007}. Total AGN plus galaxy clusters, type--2 Compton--thin AGN, type--1 Compton--thin AGN and type--2 Compton--thick 
AGN are drawn as solid lines ({\em from top to bottom}). Only the three populations together fit the observed XRB spectrum well. Compton--thin
AGN contribute most.}\label{fig:xrb}
\end{center}
\end{figure}

Recently, a new self--consistent population synthesis model for the cosmic X--ray background has been developed (\cite{Gilli2007}),
including the most recent determination of the AGN luminosity function evolution in the soft and hard X--ray bands. 
Gilli \etal~ consider a $N_{\rm H}$ distribution with a relatively small fraction of unabsorbed AGN ($ \log N_{\rm H} < 22$) but with a growing contribution at higher column densities. 
Most of the emission originates from the Compton--thin AGN population. For the spectral templates 
a more complex model is assumed: The spectral index of the sources scatters in a Gaussian fashion 
around an average value of $<\Gamma>=1.9$. Such spectral templates increase the contribution of 
Compton--thin AGN at 30 keV in the background spectrum by 30\% (as compared to $\Gamma=const$ templates). 
However, the observed background spectrum still calls for a moderatly and heavily obscured Compton--thick 
AGN population that contributes significantly around 30 keV. Moderatly obscured Compton-thick AGN have 
$24 < \log N_{\rm H} < 25$ whereas heavily obscured Compton-thick AGN have 
$\log N_{\rm H} > 25$. To date only $\simeq$40 local sources are candidates for 
Compton--thick AGN \cite{Comastri2004}.
However, the Compton--thick AGN are expected to become very numerous below current flux limits.
This fact motivates for intensive studies with future X--ray missions such as Simbol--X, Spectrum--X-Gamma, NeXT, Constellation--X 
and XEUS.

Figure~\ref{fig:xrb} shows the results from the most recent population synthesis model (Gilli \etal~2007).
The model involving three AGN populations (unabsorbed Compton--thin, absorbed Compton--thin and Compton--thick sources)
is in good agreement with the observed data points. Note that
only the inclusion of the Compton--thick population (lowest bump peaking around 20-30 keV) reproduces the total cosmic XRB 
spectrum. The fraction of Compton--thick AGN predicted by this model is consistent with the actually observed numbers in the Swift BAT, INTEGRAL and CDF--S catalogues.

\section{Conclusions}\label{sec:concl}
The major scientific topic of this spring symposium are black holes. Many theoretical contributions
have been presented at the meeting that help in understanding these mysterious objects. In contrast, 
X--ray astronomy offers the possibility to study black holes in action. Together with gravitational
waves X--rays are best suited because they allow insight into the direct surroundings of 
black holes.

This review demonstrates that X--ray astronomy is a powerful tool to understand black hole activity 
and their cosmological role in galaxy formation and galaxy evolution. Deep X--ray surveys probe
black holes at cosmological distances. It is crucial to determine the redshift of these sources to
trace back their cosmic history. This is done by optical follow--up observations with the best 
telescopes worldwide. Many X--ray surveys together provide a highly complete data set which
can be exploited scientifically.

The most relevant findings of recent AGN X--ray cosmology are the following:
\begin{enumerate}
\item Observed X--ray spectra support the standard model of AGN, that is an accreting SMBH. However,
AGN unification schemes have to consider luminosity effects and feedback mechanisms.
\item The cosmic X--ray background is a total signal of individual active galaxies and galaxy clusters.
\item An AGN classification scheme requires both optical (broad vs. narrow emission lines) and X--ray (absorbed versus unabsorbed spectrum) information.
\item The fraction of type--2 to type--1 AGN decreases with luminosity. This indicates an  
interaction between the luminous galactic nucleus and its environment. Luminous AGN seem to be 	able to clean out their environment.
\item The type--2 fraction shows a small evolution with redshift in the sense that higher redshift environments seem to have somewhat more absorption. 
\item Active galaxies do not evolve in a uniform manner. Depending on the luminosity (or black hole mass 
respectively) SMBHs exhibit an anti--hierarchical growth, i.e.\ quasars enter the cosmic stage significantly earlier 
than Seyfert galaxies. In other words: the density evolution of AGN is luminosity--dependent.
\item X--ray luminosity functions are in perfect agreement with other wavebands.
\item Compton--thin AGN contribute most to the cosmic X--ray background.
\item However, Compton--thick AGN are an essential ingredient to reproduce the total cosmic X--ray 
background spectrum. They mainly contribute at photon energies around 30 keV.
\item Accreted BHMFs from the X--ray background can be reconciled with dormant relic BHs in the local 
universe if one assumes that the BHs rotate.
\item Theory and observation indirectly support the existence of progenitors such as mini--quasars at 
$z\simeq 11$ and GRBs at $z\simeq 20$. This should be tested by direct observations with XEUS.
\end{enumerate}

X--ray astronomy is going to be successful in future if the survey capacities are extended successively.
Surveys with larger fields and deeper pencil beams are natural goals. Therefore, upcoming projects
such as eRosita aboard Spectrum--X--Gamma, Simbol--X, NeXT and XEUS are promising scientific drivers.

\begin{acknowledgments}

We thank Roberto Gilli for providing the model prediction in figure \ref{fig:evolution}.
AM acknowledges the invitation to the STScI at Baltimore. It was a great experience to discuss
current black hole research in an exciting and warm atmosphere.
\end{acknowledgments}

%
%
%

\end{document}